# An improved model of the neutrinoless double-beta decay.
## HM-results for $^{76}$Ge and a reanalysis of CUORICINO data for $^{130}$Te.

I.V.Kirpichnikov
SSC RF "Institute for Theoretical and Experimental Physics", Moscow
NRC "Kurchatov Institute"

An abstract

A model of a neutrinoless 2β-decay was proposed which predicted a shift of the 2β0ν-signal from a $Q_{\beta\beta}$ value (2β-decay energy). The shifts were calculated for $^{76}$Ge ( ΔE= –2.6 keV ), $^{100}$Mo ( ΔE= –4.7 keV ), $^{130}$Te ( ΔE= –3.7 keV ). An appearance of the shifts was explained by an excitation of atomic shells of the product atom, with the following de-excitation by emission of X-rays. A virtual excitation of the product nuclei was supposed in an addition to exchange with a virtual neutrino between the two decaying neutrons.

A comparison with the H-M and Cuoricino data has strongly supported the validity of the proposed model. It pointed out also that the 2β0ν-decay has been experimentally observed ten years ago.

Searches for the neutrinoless double beta-decay were intensively performed within several decades. Such interest could be understood as it is the only process known so far which able to prove that neutrinos have Majorana mass components, being their own anti-particle. It would demonstrate also a process which violated the lepton number by two units.

The only claim for the observation of the process was published by a group from a Heidelberg–Moscow Collaboration [1,2]. Still the recent GERDA publication [3] did not support the claim .

To understand the situation it occurred possible to correct the theoretical approach and to repeat analysis of some experimental data.

## A correction to the model of the 2β0ν-decay

Contemporary models of the 2β0ν-decay have postulated that the signal of the decay should have an energy equal the $Q_{\beta\beta}$ ( 2β-decay energy, the difference between parent and product atom masses ). Still an analysis of the experimental data for $^{76}$Ge [1-3] and $^{130}$Te [4-6] indicated a presence of possible shifts of a 2β0ν-decay signals from the proper $Q_{\beta\beta}$ values [7].

A new model of the process was proposed, which predicted a shift of the 2β0ν-signal from a $Q_{\beta\beta}$ . An appearance of the shifts was explained by an excitation of atomic shells of the product atom with the following de-excitation by emission of X-rays.

The 2β0ν-decay was followed by an energy release $Q_{\beta\beta}$ ~ several MeV. On the contrary, a reconstruction of electron shells in the process A(z,N)→A(z+2,N-2) demanded some expenditure ΔE of an energy. The consideration of the 2β0ν-decay process *had to include both the above processes,* the simultaneous energy release and expenditure, as the total 2β0ν-decay energy $Q_{\beta\beta}$ should be conserved.

First it has to be mentioned , that the signal *should be shifted relative the Q-value in any case*. A product nuclei emitted two electrons and nothing else. The new-born electrons took away some momentums, $p_1$ and $p_2$, ~ several MeV/c each one. The issue is whether the two new born electrons recoiled with the entire nucleus , or with the two new born protons. If they interacted with the entire nucleus, a proper recoil momentum got the product nucleus. The electrons lost a small part of their energies

$$\varepsilon \approx [(p_1)^2 + (p_2)^2] / 2\cdot M(Z) \quad \ldots\ldots\ldots\ldots\ldots\ldots\ldots\ldots(1)$$

where M(Z) is the mass of a product nuclei. The loss should be an order of several 0.01 keV. This energy was transferred to a kinetic energy of a product nuclei and couldn't be observed experimentally. And it was the basement of the contemporary 2β0ν-decay models.

It would be valid in the case if the momentums were accepted by the product nuclei as a whole one. But was it so? Two neutrons decayed and two pairs (a proton + an electron) were born . It was quite reasonable to suppose that the new-born electrons would share their momentums just with the new born protons . A loss of energies by electrons would be much more essential in this version of the decay :

$$\Delta E = M(Z) \cdot \varepsilon = [(p_1)^2 + (p_2)^2] / 2m(p) \quad \ldots\ldots\ldots\ldots(2)$$

where m(p) *is now the mass of a proton*. A loss of energies by electrons would be now an order of several keV.

This version of a momentums exchange seemed forbidden as there is no an excited state of nucleus for the energy to go into. But a visible contradiction with a theory could be explained by a **virtual** processes due to the total reconstruction of both the product **nucleus** and **atom shells**. One could suppose that the two new-born protons got a virtual energy which was enough to overcome both the prohibitions : the first one for a *double-beta decay* and the second for *an exchange of momentums between the new-born electrons and the new-born individual protons.* At this exchange the electrons loose a part of sum energy ΔE about several keV, which is close to the necessary energy for a reconstruction of **electron shells** of the product atom in the process A(z,n)→A(z+2,n–2).

*A virtual excitation of the product nucleus was supposed to provide this additional energy.*
The necessary excess could be estimated as a difference ΔE between sum energies of electrons in shells of a parent and a product atoms, hence ΔE being about several keV. This energy was transferred to electron shells of the product atom, thus solving a problem of an excited state absence in the new nucleus. The transfer could be effectively produced with electromagnetic forces, as electron orbits of 2β0ν-decaying atoms were partially located inside their nucleus. And this excess of energy ΔE provided an increasing of electron energies in the new born atom.

A product nucleus of the 2β0ν-decay process was born in the ground state. The proper product atom should be born in an excited state. Energies of electron shells depended on a number of protons in a nucleus. Number of protons in the product nucleus increased by two units. Energies of all electrons in the atomic shells increased accordingly. But the number of electrons in the final atom remained the same . Two new born electrons, resulting from n→p+e transition, left the system without being incorporated into any of new atomic shells (the most probable). So an additional energy ΔE was accepted by z electrons instead of z+2 ones, and several electrons were pushed to higher shells. The product atom was excited. This excitation could be removed only by emitting of X-rays, as an emission of electrons was energy disfavored.

A response of Ge and bolometric detectors for X-rays was different , so signals of the 2β0ν-decay would be different also. But in any case the signal with an energy $E_{\beta\beta} = Q_{\beta\beta} - \Delta E$ could appear in both the spectra.

The value of the ΔE is an order of several keV, since it ΔE/Q ≈ $10^{-3}$ . Such difference could be fixed only with detectors which have a proper high resolution, such as germanium or bolometric ones.

A shift of an energy of the signal $\varepsilon \approx [(p_1)^2 + (p_2)^2] / 2 \cdot M(Z)$ due to an interaction of new-born electrons with the whole nuclei could be neglected.

**A summary**

The consequences of the model application to the existing experimental data were considered in the [7]. The model removed a very serious argument against the Klapdor's claim for an observation of a 2β0ν-decay of $^{76}$Ge ( the 1.5 keV shift of the line from the $Q_{\beta\beta}$ value). The full spectrum of Gerda [3] had four events at E=2036.5 keV= $Q_{\beta\beta}$–2.6 keV in the full agreement with [2] ( the authors of [3] have attributed them to a background ). A coincidence of the predicted shift of the line ΔE= –2.6 keV with the observed value supported the Klapdor's claim.

A re-analysis of the Cuoricino data [4-6] indicated an existence of a peak at E=2023.5 keV= $Q_{\beta\beta}$ (Te)-3.5 keV [7], which could be accepted as a shifted signal of a 2β0ν-decay of $^{130}$Te. The shifted signal in a bolometric detector could be seen if a part of X-rays left the detector. And it was possible due to a $TeO_2$ crystal structure, which provided a possibility for a canalization of proper gamma-rays (see the attachment).

All results of the investigation [7] were summarized in the table I. The table contained the calculated shifts ΔE and the results of the proper experiments $\Delta E_{exper}$. It included also an expected result of a project AMoRe [8]. Still one should remember that an intensity of the shifted gamma-line and a possibility to observe it depended on a type (a structure) of the crystal which was used in the experiment.

Table I. All numbers in [keV]

| decay | $Q_{\beta\beta}$ | $E_{\beta\beta}$ | $\Delta E_{calc}$ | $\Delta E_{exp}$ | $T_{1/2}$, years |
|---|---|---|---|---|---|
| 76-Ge→76-Se enriched Ge [1,3] | 2039.0±0.005 | 2037.5±1.0(stat)±0.5(syst)[1] ≈ 2036.5 [3] | -2.6 | -1.5 -2.5 | $2.2 \cdot 10^{25}$ |
| 130-Te →130-Xe[6] bolometric detector | ≈ 2527 | ≈ 2023.5 [4+5] | -3.7 | ≈-3.5 | $1.0 \cdot 10^{24}$ |
| 100-Mo→100-Ru[8] bolometric detector | ≈ 3531 | a project | -4.7 | ??? | ??? |

*) $\Delta E_{calc}$ – calculated with $[(p_1)^2 + (p_2)^2] / 2 \cdot m(p)$   **) $\Delta E_{exp}$ – experiment

Coincidences of the predicted shifts with the results of analysis of the experimental data for $^{76}$Ge and $^{130}$Te [7] strongly supported the hypothesis under discussion.

The total signal of a bolometric detector should include both the components, the shifted and non-shifted ones. Results of an attempt to get the full shape of the 2β0ν-decay signal in the $TeO_2$ crystal were presented at the fig.1a,b. Data of [4b] and [5] were used for this purpose. A presence of the two peaks were clearly indicated (an existence of the non-shifted peak at 2527 keV was found earlier [7] but was attributed to a background)

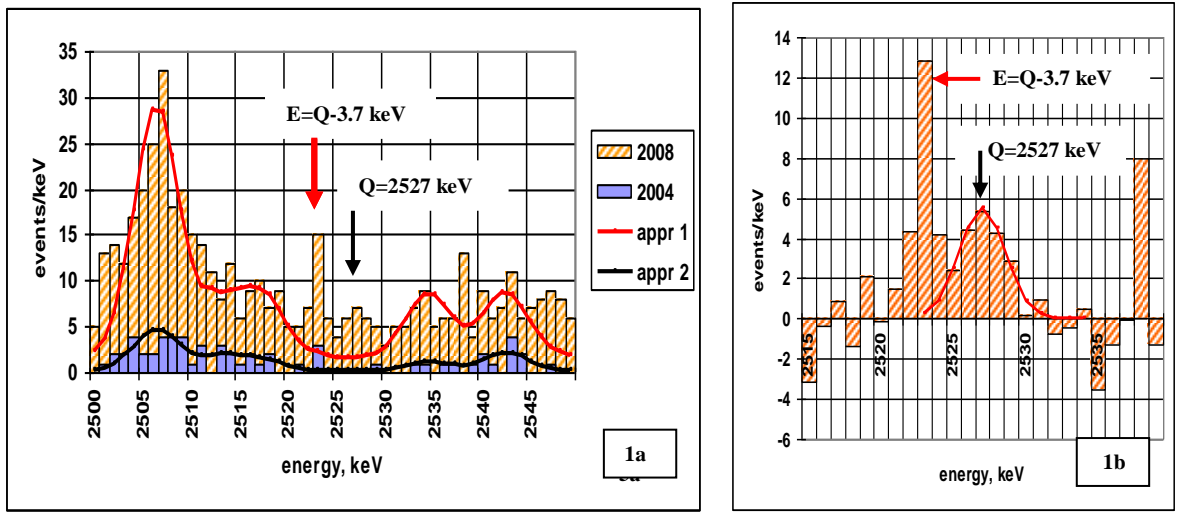

Fig.1a,b. An attempt to get a full shape of the signal .
 1a. The data [4b,5] and approximations. An existence of a background levels at 2012 , 2015, 2035 and 2043 keV was supposed. A presence of these levels was indicated in both the spectra . Energies of the levels and intensities were chosen arbitrary ones to reproduce the experimental data. The slow varying components were postulated as belonging to 2615 keV level.
 1b. A difference between the data 2008 [5] and the approximation (4 levels+1.57 events/keV). A width of 2527 keV line was about 4 keV (7.2 keV for the normal gamma-lines).
 The total intensity of the signal was
  $N_{(sign)}$ = 44.6± 8.1(stat)±4.6(sist) events  ( 2004[4b]+2008[5], Mt=17.04 kg y ).
 A corresponding $T_{1/2}=1.0 \cdot 10^{24}$ years

The analysis of [7] together with the above results (figures 1a,1b) could be taken as a proof of an experimental observation of the 2β0ν-decay of $^{130}$Te with a bolometric detector . The life-time for 2β0ν-decay of $^{130}$Te did agree with the Klapdor results [2] within an uncertainty of the matrix elements calculations .

**A conclusion**

It was pointed out definitely that the 2β0ν-decay does exist and has been experimentally observed ten years ago in the two experiments. The most seldom process was observed, which has been registered whenever in a laboratory .

It was found that two virtual processes were necessary for a realization of the neutrinoless 2β-decay. A virtual excitation of the product nuclei was postulated beyond an exchange with a virtual neutrino between the two decaying neutrons.

**Acknoledgements.**

The author was very grateful to professor F.S.Dzheparov for a fruitful discussion concerning a canalization of low-energy gamma-quanta. Just the professor F.S.Djeparov supplied the author with a scheme of a structure of a $TeO_2$ crystal.

The author was grateful to I.G.Alekseev and D.N.Svirida for an assistance in a preparation of the text for a e-print. Their help was an efficient and useful one.

**An appendix. A structure of a $TeO_2$ crystal.**

A bolometric detector fixed a local increasing of a temperature in its volume. It should be the same independently on the interaction. So the shifted signal could be seen only if a part of X-rays left the detector. And it was possible due to a $TeO_2$ crystal structure, which provided a canalization of proper gamma-rays. The figure 2 illustrated this possibility.
It presented the structure of a $TeO_2$ crystal from a wikipage [9].

$TeO_2$ viewed along a-axis:
a=0.550 nm, b=1.175 nm, c=0.559 nm. 8 units in the cell
$TeO_2$ crystal   $\lambda/2\pi = 5.33 \times 10^{-9}$ cm (e=3.7 keV)

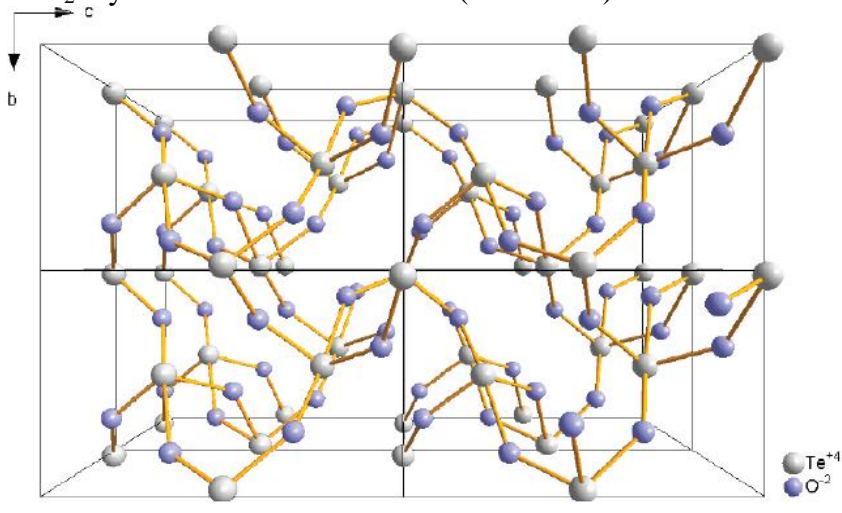

fig.2